On-chip light-scattering enhancement enables high performance single-particle tracking under conventional bright-field microscope

Pengcheng Zhang, Tingting Zhan, Guoqiang Gu, Changle Li, Mengting Lyu, Yi Zhang, and Hui Yang *

Research Center for Bionic Sensing and Intelligence, Institute of Biomedical and Health Engineering, Shenzhen Institute of Advanced Technology, Chinese Academy of Sciences, Shenzhen, China;

\*   Correspondence: hui.yang@siat.ac.cn; Tel.: +86-0755-86959041

**Abstract：**

Scattering-based single-particle tracking (S-SPT) has opened new avenues for highly sensitive label-free detection and characterization of nanoscopic objects, making it particularly attractive for various analytical applications. However, a long-standing issue hindering its widespread applicability is its high technical demands on optical systems. The most promising solution entails implementing on-chip light-scattering enhancement, but the existing field-enhancement technology fails as their highly localized field is insufficient to cover the three-dimensional trajectory of particles within the interrogation time. Here, we present a straightforward and robust on-chip microlens-based strategy for light-scattering enhancement, providing an enhancement range ten times greater than that of near-field optical techniques. These properties are attributed to the increased long-range optical fields and complex composite interactions between two closely spaced structures. Thanks to this strategy, we demonstrate that high-performance S-SPT can be achieved, for the first time, under a conventional bright-field microscope with illumination powers over 1,000 times lower than typically required. This significantly reduces the technical demands of S-SPT, representing a significant step forward in facilitating its practical application in biophotonics, biosensors, diagnostics, and other fields.

**Introduction**

Single-particle tracking (SPT) interrogates dynamic information of target objects at single-molecule/particle level, which is essential for gaining insights into the intricate mechanisms underlying a wide range of physical, chemical, and biological phenomena occurring at the nanoscale[1-3]. Implementations of SPT rely largely on the light microscopy techniques for the optical detection of individual particles, consequently enabling their localization and spatial trajectory reconstruction. Fluorescence based light microscopy provides superior background suppression to detect and locate labeled species through spectral separation. Despite its many successes in biological applications, its inherent limitations associated with labeling efficiency and fluorescence saturation restrict the amount of emitted photons and thus, the imaging speed and localization precision. On the contrary, elastic light scattering based light microscopy offers a significant advantage due to their in principle unlimited photon budget and the universal applicability of scattering phenomenon[4-7]. Driven by novel approaches to illumination, detection, and background suppression, recent advances in light-scattering based single-particle tracking (S-SPT) have led to superior or even single-molecule sensitivity at very high spatiotemporal resolution in label-free detection and tracking of nanoparticles (e.g., single dye molecules, plasmonic nanoparticles, viruses, and small proteins)[8-14]. Prominent among these are based on darkfield imaging, interferometric detection, and surface plasmon resonance microscopy. This capability makes S-SPT particularly attractive as analytical or clinical diagnostic tools for diverse purposes. However, the accessibility and widespread applicability of these S-SPT technologies are currently challenging, as they are technically demanding: they require sophisticated and expensive microscopic systems, highly stable laser sources with strong illumination power (typically 1-10 $kW/cm^2$), and in some cases meticulous image processing techniques to isolate and maximize the smallest possible scattering signals. The solution to address this issue is the implementation of on-chip light-scattering enhancement, which entails amplifying the scattering signal prior to its entry into the optical system. A range of near-field techniques have been developed to enhance the light–matter interactions through field-enhancement (e.g., via surface plasmons or optical antennas) [15, 16]. However, none of these techniques are useful for S-SPT technologies because their field-enhancement exhibits a high degree of localization, thus its enhancement range is insufficient to cover the three-dimensional trajectory of particles within the interrogation time. Additionally, they are inherently costly, requiring specific nanostructures and complex near-field configurations[17]. Direct and efficient realization of on-chip scattering enhancement beyond the confines of near-field range is highly desirable.

In this study, we present a straightforward and robust on-chip microlens-based strategy for light-scattering enhancement, providing an enhancement range ten times greater than that of near-field optical techniques. This strategy involves the use of dielectric microspheres of high refractive index (microlens chip). These symmetrical dielectric microspheres have been reported as superlenses that are capable of magnifying nearby objects before projecting them into a conventional microscope's objective lens[18-22]. Apart from that, we show that these dielectric microspheres can simultaneously enhance backscattering of visible light from individual nanoparticles ranging from near-field to semi near-field, which is ascribed to the increased long-range optical fields and complex composite interactions between two closely spaced structures[23-28]. This provides sufficient detectable photons and a high signal-to-noise ratio

(*SNR*) and enables high-performance tracking of individual plasmonic nanoparticles even under conventional bright-field microscope with incoherent light sources, as illustrated in Figure 1 (a and b). This significantly reduces the technical demands of S-SPT, representing a significant step forward in facilitating its practical application in biophotonics, biosensors, diagnostics, and other fields. As a demonstration, we study freely diffusing gold nanoparticles in solution and derive their diffusivity constant as well as hydrodynamic radius through precise tracking. We believe this demonstration will not only enable low-cost, stable and compact optical configurations, but also provide a complementary on-chip strategy for various light microscopy to further boost their performances.

**Results:**

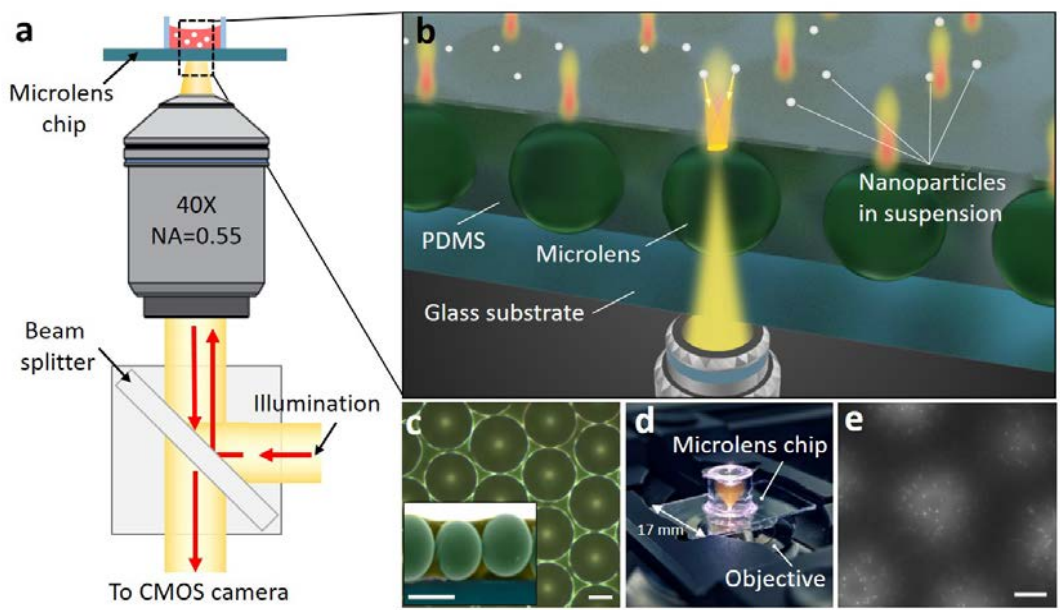

*Figure 1. Schematic illustration and experimental setup. a, Schematic of the optical*

*system. The microlens chip is positioned on a conventional inverted optical microscopy system with unpolarized incoherent light source. b, Illustration of the on-chip light-scattering enhancement. The enhanced backscattered light from individual particles is collected by the microlens and then transmitted to the objective. Note that the illustration is not drawn to scale. c, Morphology and cross-section (insert image) of the microlens chip. The images are taken by optical microscope and scanning electron microscopy (SEM), respectively. The different colors in the false colored SEM image distinguish the glass substrate, microlenses and PDMS, respectively. Scale bar = 50 μm. d, Photograph of the microlens chip under working condition. The glass substrate is clamped on the sample stage of the inverted optical microscope. A sample reservoir is made by attaching a glass ring on top of the glass substrate. e, Gray level virtual image of 100 nm AuNPs captured under microlens chip. The bright spots in the image indicate individual AuNPs. Scale bar = 20 μm.*

The microlens chip is composed of high refractive index dielectric microspheres on a glass substrate and shares rather simple fabrication process (see Methods section, Figure S1 and Figure S2). These dielectric microspheres are self-assembled into a close packed fashion and are immobilized by a thin layer of Polydimethylsiloxane (PDMS), as shown in Figure 1c. Each microsphere acts as an individual microlens that generates magnified virtual image and projects to the objective lens of the optical system. In accordance with previous reports, the imaging characteristics of microlenses, such as image position, lateral magnification, and field of view (FOV), have been found to be associated with parameters including microlens size, refractive index (RI), and the RI ratio of the microlens to its surrounding medium. This provides the flexibility to fine-tune the imaging properties of microlens chips through varying combinations of these parameters. In our current study, as a demonstration, we utilized commercially available dielectric microspheres with diameters of 73.5 μm (±2.0%) and RI of 1.92.

The optical system is based on a conventional inverted optical microscope equipped with a ×40 low numerical aperture (NA) objective lens (NA = 0.55, ZEISS LD A-Plan) and illuminated under unpolarized incoherent light source (white LED light source, ZEISS Colibri) as illustrated in Figure 1a. The microlens chip is positioned above the objective and nanoparticle solution is dropped onto the chip (Figure 1d). Here in the reflection-based illumination mode, the backscattered light of nanoparticles is collected by the microlenses, captured by objective and finally imaged onto a CMOS camera, generating a magnified virtual image (the same orientation as the objects) of nanoparticles in the far field. As a point-like scatterer, individual nanoparticles appear as a diffraction pattern on the image plane, known as the point spread function (PSF). We first immobilized the gold nanoparticles (AuNPs) on the surface of the microlens and imaged them under unpolarized white light source. Typical image of AuNPs with 100 nm in diameter observed with microlens chip is shown in Figure 1e. With a ×40 objective lens, each microlens generates a magnified virtual image where the single AuNPs appear as bright spot-like PSF against the background in the field of view (FOV≈25 μm in diameter in the real space). Consistent with the characteristics in microsphere-assisted imaging, the central region of the FOV exhibits superior image quality with optimal brightness and minimal distortion. Towards the edges of the FOV, the brightness of their corresponding PSFs decreases, and their circular shape distorts into a spindle-like shape due to the distortion. In our study, we define region of interests (ROIs) within the FOV characterized by a distortion rate of less than 1% (more details to the distortion rate can be found in Supplementary Text and Figure S3). This corresponds to an area of approximately 6.9 μm in diameter within the real space imaged by each microlenses. Accordingly, only the AuNPs located within this area were subjected to analysis. Within this region, the variation in PSF brightness is shown to be less than 9% (Figure S4). Further details regarding the distortion rate will be presented later.

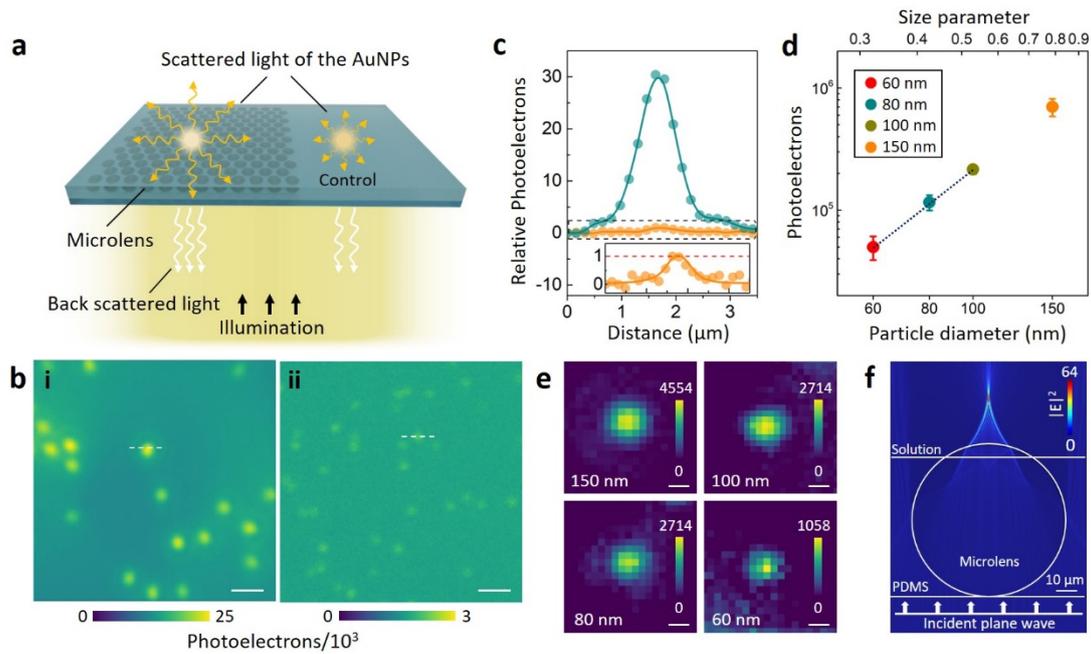

*Figure 2. Enhancement effect of the microlens chip on the light-scattering from nanoparticles. a, Schematic illustration of the experimental configuration designed for on-chip light-scattering investigation. Back scattered light from the AuNPs on the substrate with and without microlenses (control sample) are collected respectively. b, Photoelectron images of 150 nm diameter AuNPs imaged with (i) and without (ii) the use of microlens chip. Photoelectron data were acquired at exposure time of 1.5 ms with white light illumination (power intensity = 0.7 W/cm$^2$). Note that for the convenience of visual identification, the scale bars of the photoelectron are different. Scale bar = 2 μm. c, Comparison of their relative number of photoelectrons from individual AuNPs indicated by dashed lines in (b). Green and orange dots and lines denote the AuNP imaged with and without the microlens chip, respectively. Their photoelectrons are normalized by the maximum value of AuNP imaged without the microlens chip. Insert: the enlarged view of the AuNP imaged without the microlens chip. d, Measured photoelectrons from AuNPs of different sizes on the use of microlens chip.*

*Photoelectron data were acquired at exposure time of 1.5 ms with white light illumination (power intensity = 0.7 W/cm$^2$). The size parameter is defined as α = πd/λ, where d is the diameter of the particle, and λ is the wavelength of the incident radiation (λ = 600 nm). The dashed line represents the linear fitting of the three values, with slope of 2.87. This indicates approximately a d$^3$ diameter dependence of the photoelectrons under the microlens chip. e, Relative photoelectron images of AuNPs of different diameter (150 nm, 100 nm, 80 nm, and 60 nm). Photoelectron data were acquired at exposure time of 200 μs with white light illumination (power intensity = 1 W/cm$^2$). The photoelectron value in each image is obtained by subtracting their corresponding background value. Scale bar = 50 nm. f, Finite element method simulation of the electric field (|E|$^2$) distribution generated by the microlens. The diameter of the microlens is 73.5 μm and the wavelength of the illumination light is set as 600 nm. The simulation shows that the incident light is converged by the microlens within its FOV at its shadow-side. Further details refer also Supplementary Figure S6.*

To demonstrate the light-scattering enhancement afforded by microlenses, we compare bright-field microscopy images of 150 nm diameter AuNPs imaged with and without the use of microlens chip under the same illumination power (Figure 2a). The diameter was chosen to ensure a sufficiently intense light-scattering signal while observed directly under ×40 objective lens without microlenses. Surface immobilized AuNPs immersed in water are illuminated with unpolarized white light, and light scattered by individual nanoparticles is collected. To compare their photoelectron counts, the gray level images produced by the CMOS camera are converted and presented in terms of photoelectrons (details can be found in Methods section).

As expected, AuNPs imaged with microlens exhibit higher scattering intensity compared to those imaged without the microlens. While bright visible intensity was captured for the AuNPs imaged with microlens, only inconspicuous intensity was observed for the control samples, as shown in Figure 2b. To quantify the effect of scattering enhancement, the enhancement factor $\delta$ is defined as $\delta=N_{PE\text{-}\mu lens}/N_{PE}$, where $N_{PE\text{-}\mu lens}$ and $N_{PE}$ denotes the integration number of net photoelectrons with and without microlens, respectively. It is shown that the microlens chip largely elevates the number of photoelectrons from individual 150 nm AuNPs, leading to an enhancement factor of approximately 76 under the same illumination power (0.7 W/cm$^2$), as shown in Figure 2c. Notably, this considerable boost in the scattered photons is not accompanied by a corresponding increase in background noise, resulting in a significant enhancement of ~16-fold in the *SNR*. Note that no contrast enhancement or average has been performed on any of these images, except for the autogain of the camera.

We next assess the performance of microlens chip on smaller AuNPs. Previous studies including simulations and experimental measurements have revealed that, already for AuNPs smaller than 100 nm, the scattering cross-section of individual AuNPs exhibits a rapid decay with its size[29, 30].

A challenge thus arises when to image small nanoparticles, as a very limited number of scattered photons (proportional to its scattering cross-section) are expected. AuNPs with diameter of 100 nm, 80 nm, and 60 nm are investigated, respectively. Without the microlens chip, reliable detection of individual AuNPs under a conventional bright-field microscope becomes challenging, and determining their corresponding enhancement factor is not feasible. However, these AuNPs can be clearly observed upon utilization of the microlens chip. We analyzed the dependency of photoelectrons on their sizes under microlenses, as shown in Figure 2d. In contrast to the anticipated rapid decay in scattering intensity, we observed a much slower decay of scattering intensity from these AuNPs. For instance, despite having a scattering cross-section that is ~21 times smaller than that of 100 nm AuNPs, a 60 nm AuNP (scattering cross-section ~700 $nm^2$ calculated at 633nm)[29] exhibits only a ~3 times reduction in the amount of backscattered photons. The highly increased abundance of photons allows us to capture surface immobilized single AuNPs with submillisecond temporal resolution while maintaining sufficient *SNR* under low power illumination (Figure 2e). For instance, 60 nm AuNPs can be captured at exposure time of 200 μs (corresponding to a frame rate of 5000 Hz, limited by the camera readout time) at illumination power density of 1 W/$cm^2$ with *SNR* > 20. As the *SNR* is directly proportional to the illumination power, it is possible to achieve further enhancements in *SNR*, as well as improvements in exposure time by using higher power illumination sources (Figure S5). To the best of our knowledge, this is the first time that such small AuNPs can be captured with such high temporal resolution under bright-field white-light microscope. This enhancement on backscattered photons of AuNPs in our microlens chip arises from the enhanced backscattering effect featured in the system consist of nanoparticle and dielectric microsphere[23, 24, 31, 32]. This effect is attributed to three distinct factors: i) the intensified illumination intensity achieved through the implementation of microlenses (Figure 2f), ii) the improved collection efficiency resulting from the enhanced NA (Figure S6), and iii) the mutual interactions between AuNPs and microlenses. This is in significant contrast to commonly used scattering-based imaging approaches such as darkfield, interferometric or plasmonic imaging, where scattering enhancement relies largely on the increase of illumination power.

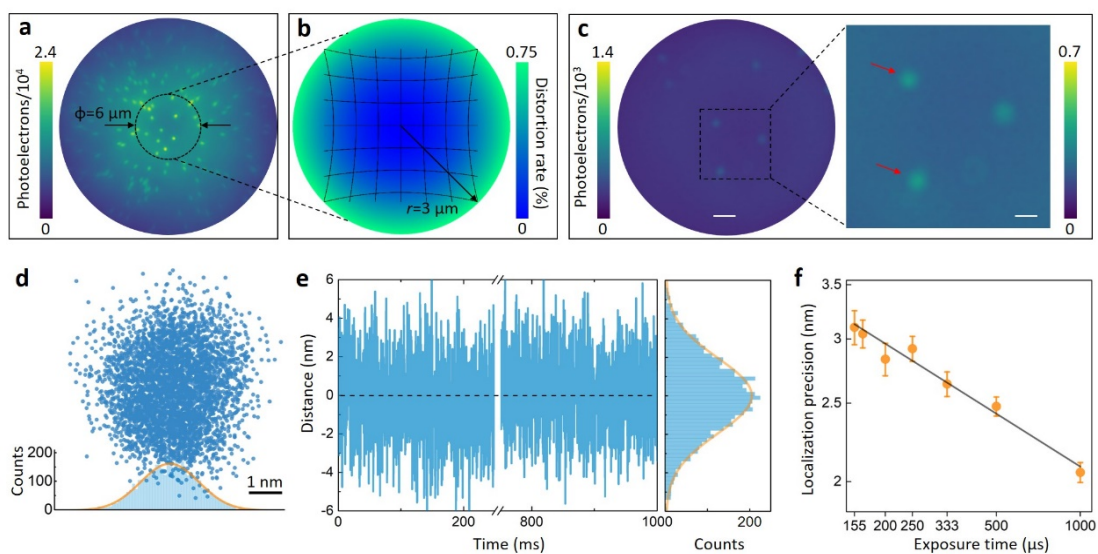

*Figure 3. Determination of the localization precision under microlens chip. a, Photoelectron images of AuNPs (100nm) in the FOV zone imaged with microlens chip. Photoelectron data were acquired at exposure time of 15 ms with white light illumination (power intensity = 0.5 W/cm$^2$). b, Calculation results of distortion rate in the corresponding area (black dashed circle) from (a). c, Photoelectron images of immobilized AuNPs (60nm). The red arrows indicate the adjacent AuNPs used to assess the localization precision. Scale bar = 20 μm (left image) and 10 μm (right image), respectively. d, Scatter plots of a 60 nm immobilized AuNP imaged at 200 μs exposure time with white light illumination (power intensity = 1 W/cm$^2$). The lower image shows the histogram of its x coordinates. e, Distance fluctuations of the adjacent AuNPs from the analysis of the image sequences in (c). The distance fluctuations were obtained by subtracting the mean distance value from the actual distance value. The corresponding histogram is on the right. f, Experimental localization precision (orange dots) of the 60 nm AuNPs and theoretical shot-noise limited behavior (black line) measured by varying the exposure time with white light illumination (power intensity = 1 W/cm$^2$).*

After the immobilized AuNPs can be imaged with high temporal resolution via on-chip light-scattering enhancement, we extract their positions and investigate the achievable localization precision under different exposure time. Localization precision refers to the accuracy with which the positions of individual nanoparticles can be determined from experimental data. In a shot noise limited system, the localization precision is inversely proportional to the square root of the photon number scattered by the nanoparticle[8]. Consequently, having a surplus of collected photons can not only lead to the increase in temporal resolution, but also the increase in spatial resolution. The position $(x, y)$ of individual nanoparticles can be obtained by the equation $(x, y) = (x', y')/\beta_{lat}$. Here, $(x', y')$ is the nanoparticle position extracted from the magnified virtual image via single-particle tracking software and $\beta_{lat}$ denotes the lateral magnification of the microlens on the nanoparticles (details on the single particle tracking and localization can be found in Supplementary Text). However, the microlenses, similar to traditional spherical lenses, have a varying lateral magnification that changes with the position of the objects away from the optical axis in the *x-y* plane. In particular, a gradual increase in lateral magnification with the distance from the optical axis, known as pincushion distortion, occurs in microsphere assisted imaging as mentioned before [33, 34]. As can be seen form Figure 3a, while individual nanoparticles are located further away from the optical axis, their PSFs changes from circular to spindle-shaped, indicating a change in lateral magnification. To precisely locate the positions of nanoparticles within the FOV, it is necessary to understand the impact of pincushion distortion on lateral magnification. We define the distortion rate $\eta = [(\beta_1 - \beta)/\beta_1] \times 100\%$, where $\beta_1$ is the actual lateral magnification, and $\beta$ is the ideal lateral magnification at optical axis. Geometric optics can be used to calculate the pincushion distortion rate of the microlenses with tens of micrometers

(details on the calculations can be found in Supplementary Text) [35]. According to the calculations, the pincushion distortion rate is relatively small at the center of the FOV (e.g., in the area of $\Phi=6$ μm in the FOV, $\eta<0.8\%$, Figure 3b), but it increases non-linearly (approximate to polynomial growth) from the optical axis to the edge of the FOV (Supplementary Figure S3). It is thus possible to margin an appropriate ROI in the FOV with a known level of error. In our experiments, for every virtual image generated by individual microlenses, ROI is defined as the area where the distortion rate is lower than 1% (e.g., corresponding to an area of $\phi=6.9$ μm). This introduces negligible position error both in the x and y directions ($\delta_{x,y}<1\%$). For specific applications that require large-scale observation area, post-image processing based on distortion correction can be used to further extend the usable range of ROI[34]. Apart from pincushion distortion, the lateral magnification of nanoparticles is also influenced by their vertical position along the *z*-axis within the depth of field (DOF). However, this effect is also relatively minor, with an impact of approximately $\delta_z<1.1\%$ within the DOF based on calculations (Figure S3). Therefore, in the ROI, the lateral magnification of the nanoparticles can be regarded as constant for simplicity, with the introduced overall position error $\delta_r<2\%$. Accordingly, only the nanoparticles located within ROI were subjected to analysis.

In our experiments, to avoid mistaking sample vibrations induced localization error, we evaluate localization precision by quantifying immobilized interparticle distances and express it in terms of the standard deviation (SD) of the measured distances (Figure 3c). An example of scatter plots of an individual 60 nm immobilized AuNP imaged at 200 μs exposure time is depicted in Figure 3d. Thanks to the light-scattering enhancement afforded by microlenses, we show that a single particle localization precision of 2.9 nm ($\sqrt{2}\times$SD, SD=2.05 nm) at 200 μs exposure time (5000 Hz) under bright-field microscope can be achieved (Figure 3e). This is on par with state-of-the-art nanoparticle localization techniques on such short timescales, but now this is achieved with much simple optical configuration at conventional optical microscopy system at illumination power (1 W/cm$^2$) that are >1,000-fold lower than typically required value. The plot on localization precision versus number of photoelectrons at different exposure time further confirms the relationship between shot noise limited and experimentally determined localization precision (Figure 3f).

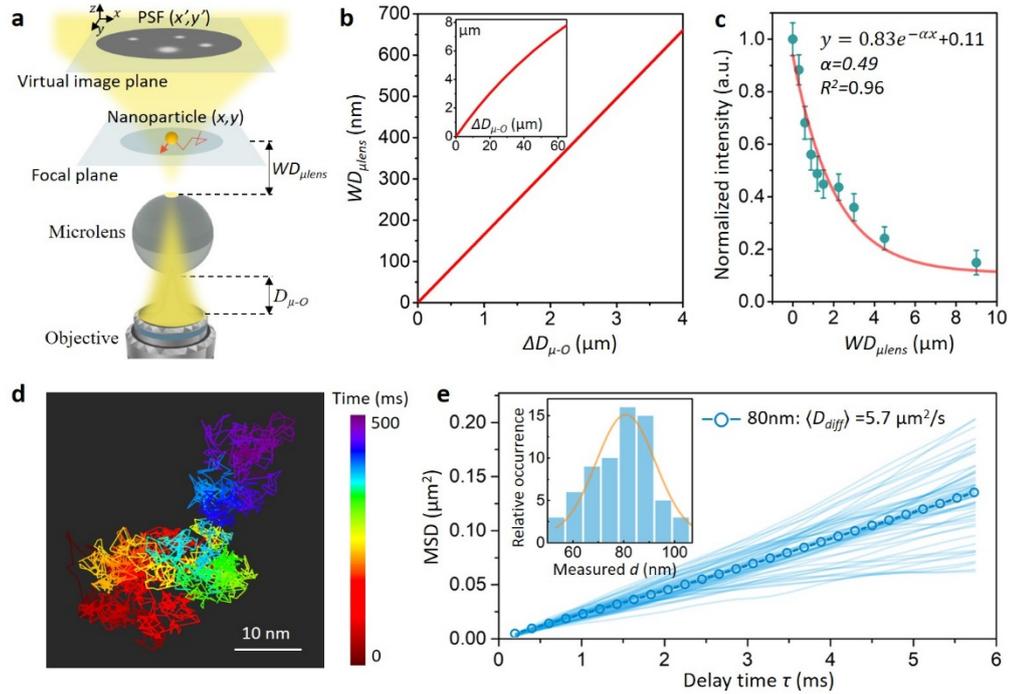

*Figure 4. Tracking the moving nanoparticles in the solution. a, Illustration of the formation of the virtual image. The microlens acts as an auxiliary lens that forms a magnified virtual image above the specimen's surface that is captured by the objective lens. b, The calculated relationship between the displacement of the objective ($\Delta D_{\mu\text{-}O}$) and the working distance of the microlens ($WD_{\mu lens}$). The diameter of the microlens is 73.5μm and the working distance of the objective is 2.3 mm as indicated by the manufacturer. Insert: the calculated $WD_{\mu lens}$ curve under the $\Delta D_{\mu\text{-}O}$ at the range of 0-65 μm. c, Normalized intensity of the AuNPs versus the working distance of the microlens. An exponential fits give the decay constant α = 0.49 with a squared correlation coefficient ($R^2$) of 0.96. d, Trajectory of a single AuNP undergoing Brownian motion recorded over 500 ms at 200 μs exposure time. e, MSD(τ) curves extracted from individual trajectory of 80nm AuNPs. The thin lines show MSD data from individual trajectories that contained at least 28 trajectory points. Thick lines with*

*circles show their weighted average. Diffusion constant extracted from the fits are listed in the legend. Insert: histogram of AuNP diameters extracted from the SE relation.*

Having a method for capturing the immobilized AuNPs with high spatiotemporal resolution with microlens chip, we set out to capture the AuNPs that move in suspension. AuNP solution (dissolved in water, ~5× $10^9$ particle/ml) is dropped on the top surface of the microlens chip. As has been shown in numerous studies on microsphere assisted imaging, under the virtual imaging mode, focal plane of the microlens, corresponding to its working distance $WD_{\mu lens}$, can be adjusted by altering the distance $D_{\mu\text{-}O}$ between the microlens and the objective, thereby enabling the observation of the objects located at different positions along the *z*-axis direction (Figure 4a). To observe the movement of nanoparticles in solution, we adjusted the $WD_{\mu lens}$ upwards by reducing the distance $D_{\mu\text{-}O}$. Note that, due to the hydrodynamic drag induced wall effect, the movement of single nanoparticles can be different in velocities depending on their distance *h* from the plane wall (i.e., top surface of the microlens). As the distance-to-diameter ratio (*h/d*) exceeds the critical value of 12, the impact of the wall effect is shown to be negligible[36]. Therefore, depending on specific investigation, the focal plane of the microlens should be determined so as to target appropriate layers of nanoparticle along the *z*-axis direction. Calculations based on FEM and ray optics were performed to more rigorously determine how distance $D_{\mu\text{-}O}$ affect the $WD_{\mu lens}$ of the microlens

(details on the calculation method can be found in Supplementary Text). It is shown that the $WD_{\mu lens}$ can be levitated gradually as the objective approaches to the microlens. For example, when the objective approaches to the microlens by 1 µm, the focal plane of the microlens undergoes a levitation of approximately 150 nm, corresponding to a conversion ratio ($R_{conv}$) of 0.15 (150 nm/µm), as shown in Figure 4b. Calculation results also show that a lower conversion ratio, i.e., < 0.1, can be achieved by using microlenses of smaller size (Figure S7). It is noted that, as the focal plane is levitated from the surface of the microlens, the targeted AuNPs in the FOV are subjected to have a decrease in intensity and thus SNR, primarily due to the reduction in the collection efficiency of the microlens on their backscattered photons[26, 37]. Experimental measurements on the scattering intensity of AuNPs at different focal plane showed a nearly exponential decay of the intensity with the $WD_{\mu lens}$ (Figure 4c). We define the characteristic working distance $WD_c$ as $WD_c = 1/\alpha$, where $\alpha$ represents the decay constant obtained from the exponential fitting of the experimental data in Figure 4c. For our current settings, we observe a much slow decay constant ($\alpha = 0.00049$ nm$^{-1}$) thus a long decay length, which results in a $WD_c$ of approximately 2.0 µm. Compared to the commonly used total internal reflection (TIR) based near-field imaging approaches, which typically have a characteristic working distance of ~200 nm (limited by the penetration depth of evanescent wave), our approach offers a working range that is 10 times larger, enabling the possibilities for investigations beyond surface studies. Note that, owning to the highly increased abundant of photons, $WD_{\mu lens}$ as large as 6 µm can still be applicable without losing the advantage of high *SNR* (*SNR* > 20). As the working distance is increased, there is a slightly variation on the distortion rate (Figure S3). Nevertheless, the induced position error can be ignored.

We are able to track individual moving AuNPs as small as 60 nm with high spatiotemporal resolution under bright-field microscope. To further test the feasibility of our method, we track AuNPs undergoing Brownian motion and deduce their diffusivity as well as the nanoparticle size from their trajectories. In our experiments, we adjust the focal plane of the microlens with enough distance (*h/d* > *12*) from the substrate to diminish the wall effect as well as to ensure sufficient *SNR*. The diffusion of a particle in a fluid is described by the Stokes–Einstein (SE) equation and one can arrive at its hydrodynamic diameter ($d_{hydr}$) by evaluating their diffusivity ($D_{diff}$) from the mean squared displacement (*MSD*) of a particle trajectory (details can be found in Supplementary Text). To measure diffusivity $D_{diff}$ with higher precision, more trajectory points and thus fast recordings are highly desirable. However, high-speed imaging requires a large *SNR* to ensure low localization error. The ability to maintain a high *SNR* even at high-speed imaging makes our approach an ideal technique for precise nanoparticle tracking. As an example, nanoparticles of 80 nm were tracked at 200 µs time resolution (5000 Hz). Under a specific recording time (limited by the memory buffer of the camera and the computer), the number of captured trajectories is related to the particle concentration. Meanwhile, the number of localization points in each trajectory are limited predominantly by the axial diffusion of nanoparticles. For a particle concentration of ~5×10$^9$ particle/ml, it is possible to record more than 20 trajectories using 1-second-long videos in the ROI of a single microlens, with each trajectory consisting of over 100 localization points. An example image sequences of 7 consecutive frames are shown in Supplementary Figure S8. By extraction their position in *x-y* plane on each image, the trajectories of individual nanoparticles (AuNPs of 80 nm) can be obtained (Figure 4d). From each trajectory, the mean squared displacement curves ($MSD(\tau)$) can be obtained (thin lines in Figure 4e). The thick curve in Figure 4e shows the resulting averaged MSD plots (n > 50) based on a linear fit of the MSD data, which confirms the free diffusion from its linear relationship. As an example, for the AuNPs of 80 nm, an excellent agreement with the theoretical value of $\langle D_{diff} \rangle = (5.7 \pm 0.3)$ µm$^2$/s and a deduced hydrodynamic radius of $d_{hydr} = 81.6$ nm was obtained.

**Discussion:**

In summary, we present a straightforward and robust on-chip microlens-based strategy for light-scattering enhancement, which enables high-performance S-SPT under conventional bright-field microscope at illumination powers that is >1,000-fold lower than typically required value. Besides, we show that the microlens chip provides an enhancement range ten times greater than that of near-field optical techniques, thereby enabling investigations beyond surface studies. This significant enhancement on elastic light-scattering is ascribed to the enhanced long-range optical fields and complex composite interactions between two closely spaced structures. From a fundamental perspective, this symmetrical dielectric microstructure displayed a previously less explored microsphere-nanoparticle system that combines imaging capabilities with outstanding light-scattering enhancement from the near-field to semi near-field range.

Although this study demonstrates a proof-of-concept single-particle tracking with conventional bright-field microscope, this on-chip concept can be extended to diverse microscopy techniques, thereby enhancing their overall performance at virtually no extra cost. We believe this strategy offers inspiring and ideal solutions to three key aspects: 1) reducing the reliance on complex optical systems for scattering detection, 2) enhancing the performance of existing analytical tools based on optical measurement, and 3) providing optical imaging techniques for research beyond surface study.

Our microlens chip uses dielectric microspheres, which are cheap and already commercially available for mass production. The direct fabrication of microsphere arrays over large areas takes advantage of simple solution processes routinely used in current self-assembly methods. Further developments of such microlens based on-chip strategy could aim to meet the demands of applications in biophotonics, biosensors, and diagnostics that require compact yet high-performance properties.

**Materials and Methods**

**Materials and regents**

High refraction index glass microspheres were purchased from Cospheric Co., Santa Barbara, United States. Polydimethylsiloxane (PDMS, SYLGARD 184) was purchased from DOW CORNING, United States. Gold nanoparticles were purchased from BBI solutions, Caerphilly, United Kingdom. (3-Aminopropyl) triethoxysilane (APTES) was purchased from Sigma-Aldrich, China. Milli-Q water was used for all experiments. All the regents were used as received.

**Fabrication of the microlens chip**

The fabrication process of the microlens chip is illustrated in Fig. S1. Briefly, a glass substrate (thickness 1 mm) was cleaned with ethanol, rinsed with deionized water, and dried in a stream of nitrogen. 30 μL suspension of the high refractive index glass microsphere (size ~73.5 μm, dissolved in water) was dropped onto the glass substrate, and subsequently dried at 70 ℃ for 15 min. After the suspension was completed dried, a layer of PDMS (mixing ratio 10:1) was spin-coated on the glass substrate. The thickness of the PDMS layer was adjusted without submerging the microspheres to ensure the targets are within the working distance of the microlenses (Figure S2). This can be controlled via tuning the coating speed and coating time. In our experiments, the spin coating was conducted at 5000 rpm for 5 min. The microlens chip was then de-bubbled and cured at 70 ℃ for 1 hour to cure the PDMS layer. After curing the PDMS, a glass ring is positioned on top of the microlens chip to hold the solution. In some experiments, further treatments was conducted to generate positive surface charges on the microlens chip. For this purpose, the microlens chip was treated with air plasma (30 W, 2 min) and further immersed in 5% APTES solution at 60 ℃ water bath for 30 min. This generates electrostatic attraction to the negatively charged gold nanoparticles (AuNPs) in the solution.

**Imaging the nanoparticles with microlens chip and conversion of image in terms of photoelectrons**

For the imaging experiments, AuNPs are used as an example. 30 μL AuNPs solution were dropped into the sample holder of the microlens chip and observed under inverted optical microscope equipped with visible LED light source (ZEISS Axio Observer). The backscattering signal was collected by a 40× objective (NA = 0.55) and finally transmitted to a scientific CMOS camera (Hamamatsu C13440). The objective was focused on the virtual imaging plane of the microlens. The exposure time was set between 166 μs-20 ms depending on the requirements and the region of interest (ROI) was adjusted accordingly. For exposure time below 300 μs, a small ROI (typically 512×32 pixels, limited by the reading out speed of the hardware) was used. Only the AuNPs located in the field-of-view zone were included in the statistics. The intensity of the AuNPs is represented by the gray level values ($Singnal_{Graylevel}$) measured from the images. To compare their photoelectron counts ($N_{PE}$), the gray level images produced by the CMOS camera are converted and presented in terms of photoelectrons by $N_{PE}=(Singnal_{Graylevel} - Bias) \times Gain$, where Bias is 100 gray level and Gain is 0.46 photoelectron/gray level for our CMOS camera under 16 bit-depth. The detected photons ($N_p$) can be obtained by $N_P=N_{PE}/QE$, where $QE$ stands for the quantum efficiency of the CMOS camera. Note that, $QE$ varies with wavelength of the photons. As a general comparison, a typical value of $QE=82\%$ at the central wavelength of 600 nm in our camera can be used.

## ASSOCIATED CONTENT

**Data availability:** The data that support this study are available from the corresponding authors upon reasonable request.

**Acknowledgment:** We acknowledge financial support from the National Natural Science Foundation of China (NSFC, No. 62205366, 62074155, 62175252), Shenzhen Science and Technology Innovation Commission (JCYJ20210324101405016).

**Author contributions:** H.Y. and P.Z. developed the concept. P.Z.

developed the methodology. P.Z. and T.Z. performed the experiments with assistance from C.L., M.L., and Y.Z.. G.G. performed FEM simulation. P. Z. wrote the original draft and H.Y. reviewed and edited. All authors discussed the results.

**Conflict of interest:** The authors declare no competing interests.